\documentclass[12pt,preprint]{aastex}

\newcommand{\msol} {M$_{\odot}$}
\def\lesssim{\mathrel{\hbox{\rlap{\hbox{\lower4pt\hbox{$\sim$}}}\hbox{$<$}}}}
\def\gtrsim{\mathrel{\hbox{\rlap{\hbox{\lower4pt\hbox{$\sim$}}}\hbox{$>$}}}}

\slugcomment{Accepted for publication in ApJ letters}

\shorttitle{The progenitor of SN~2002ap}
\shortauthors{Smartt et al.}

\begin{document}


\title{On the progenitor of the Type Ic supernova 2002ap}


\author{S.J. Smartt\altaffilmark{1}, P.M. Vreeswijk\altaffilmark{2}, 
E. Ramirez-Ruiz\altaffilmark{1}, 
G.F. Gilmore\altaffilmark{1}, 
W.P.S. Meikle\altaffilmark{3}, 
A.M.N. Ferguson\altaffilmark{4,7} 
J.H. Knapen\altaffilmark{5,6}}

\altaffiltext{1}{Institute of Astronomy, University of Cambridge, 
Madingley Road, CB3 OHA, Cambridge, England}
\altaffiltext{2}{Astronomical Institute ``Anton Pannekoek'', University of 
Amsterdam \& Center for High Energy Astrophysics, Kruislaan 403, 
1098 SJ Amsterdam, The Netherlands}
\altaffiltext{3}{Astrophysics Group, Blackett Laboratory,
Imperial College, Prince Consort Road, London SW7 2BZ, England}
\altaffiltext{4}{Kapteyn Astronomical Institute, University of Groningen, 
P.O. Box 800, 9700 AV Groningen, The Netherlands}
\altaffiltext{5}{Isaac Newton Group of Telescopes, 
Apartado 321, E-38700 Santa Cruz de La Palma, Spain}
\altaffiltext{6}{University of Hertfordshire, Department of 
Physical Sciences, Hatfield, Herts AL10 9AB, England }
\altaffiltext{7}{
Visiting Astronomer, Kitt Peak National Observatory, National Optical
Astronomy Observatory, which is operated by the Association of
Universities for Research in Astronomy, Inc. (AURA) under cooperative
agreement with the National Science Foundation.}


\begin{abstract}

This letter presents wide-field optical and and NIR ($UBVRI {\rm H}\alpha
K$) images of the galaxy M74 which were taken between 0.6$-$8.3 years
before the discovery of the Type\,Ic supernova~2002ap.  We have
located the position of the supernova on these images with an accuracy
of $0.3''$.  We find no sign of a progenitor object on any of the
images.  The deepest of these images is the $B-$band exposure which
has a sensitivity limit corresponding to an absolute magnitude of $M_B
\leq -6.3$.  From our observed limits, we rule out as the
progenitor all evolved states of {\em single} stars with initial
masses greater than 20\msol\ {\em unless} the WR phase has been
entered.  Two popular theories for the origin of Type\,Ic supernovae
are the core collapse of massive stars when they are in the WR phase,
or the core collapse of a massive star in an interacting binary which
has had its envelope stripped through mass transfer.  Our 
prediscovery images would be sensitive only to the most luminous 
$\sim$30\% of WR stars, hence leaving a substantial fraction of 
typical WR stars as viable progenitors. 
The energetics measured from modelling the initial
lightcurve and spectral evolution 
of SN~2002ap suggest an explosion of a 5\msol\ C+O core. While 
WR stars generally have measured final masses greater than this, the 
uncertainties associated with the explosion model, stellar evolutionary 
calculations and mass measurements suggest we cannot definitively rule out a 
WR star progenitor. The alternative scenario is that the
progenitor was a star of initial mass
$\sim20-25$\msol\ which was part of an interacting binary and stripped
of its hydrogen and helium envelope via mass transfer.  We discuss
future observations of the supernova environment which will provide
further constraints on the nature of the progenitor star.

\end{abstract}


~\clearpage


\section{Introduction}

Supernova~2002ap was discovered by Yoji Hirose on 2002 January 29.4 UT
in the spiral galaxy M74 \citep{nakano02}.  It was discovered
at $V=14.54$, and at a distance of approximately 7.3\,Mpc, may
be the closest supernova since SN~1993J in M81 (at 3.6\,Mpc). Several
observers rapidly obtained spectra, and reported that it appeared
similar to the peculiar SN~1998bw and caught at an
earlier epoch \citep[e.g.][]{meikle02}. Later optical spectra of
SN~2002ap indicate that it does appear to be a Type Ic, 
and its optical lightcurve appears to
have peaked at approximately $M_V \simeq -17.5$, some 1.7$^{m}$
fainter than SN~1998bw. Unlike SN~1998bw, there has been no
detection of a $\gamma$-ray burst (GRB) which could in any way be
coincident with the position of SN~2002ap \citep{hurley02}.
However \citet{yam2002} suggest that their more accurate determination of 
the date of peak luminosity means the time-frame for which gamma-ray data
should be searched needs to be extended, and this has not yet been 
done systematically. In a
preliminary spectral analysis \citet{mazz2002} suggest that it had a
kinetic energy $\sim4-10\times10^{51}$\,ergs, a factor of roughly 10
less than that of SN~1998bw, but similar to the hypernova 
SN1997ef \citep{iwamoto2000}.  The spectral
similarity to SN~1998bw, the possible link between very energetic
supernovae and GRBs, and the lack of substantive data on rare Type\,Ic
events make this bright supernova a very important object to monitor
and study in detail.

A distance of 7.3\,Mpc ($\mu_0=29.3$) to the galaxy M74 (= NGC~628)
has been determined by \citet{sharina96} and 
\citet{sohn96} from the magnitudes of the
brightest blue and red supergiants. Although this method suffers from
significant uncertainties \citep[typically $\pm0.5^{m}$ in
$\mu_0$;][]{rozanski94}, M74 is certainly close enough to allow
extensive observations of this bright supernova for some time.
Supernova Types~II, Ib and Ic are thought to originate in the collapse
of the iron cores of massive stars at the end of their nuclear burning
lifetimes. But the types of stars that cause these, are not well
constrained by observations. The only definite detection and
determination of the spectral type of a supernova progenitor is that
of Sk$-69^{\circ}202$, the precursor to SN~1987A \citep[a B3Ia
supergiant;][]{wal89}.  The progenitor of SN~1993J in M81 was
identified as a possible K0\,Ia with some excess $UB$ band flux from
either an unresolved OB association or a hot companion
\citep{alder94}. In two recent papers Smartt et al. (2001, 2002) 
presented high-resolution prediscovery images of
two nearby Type\,II-P supernovae, SNe~1999gi and 1999em.  In neither
case was an actual progenitor star detected despite the depth of the
exposures, however upper mass limits of the progenitor stars were
derived of 
9$^{+3}_{-2}$M$_{\odot}$, and 12$^{+1}_{-1}$M$_{\odot}$ respectively.
Type\,Ic supernovae probably arise in stars which
have lost their H and He envelopes, and possible candidates are
massive WR stars, or stripped high- or intermediate-mass stars in
interacting binaries. In this {\em Letter} we present analysis of optical
and near-IR images of M74 taken before the explosion of SN~2002ap, and
ascertain if there is any sign of a massive luminous progenitor star.

\section{Observational data and analysis}

We have 2 sets of wide-field optical images of M74 taken before
discovery of 2002ap from different sources. The first set are from
the Wide-Field-Camera (WFC) on the Isaac Newton Telescope (INT), La Palma,
taken on 2001 July 24 through filters $UBVI$. The exposures were 120s
in each of $BVI$ and 180s in $U$. These were taken at the end of a night
during the Wide Field Survey programme on Faint Sky Variability
\citep{groot02}. The WFC comprises 4 thinned EEV 4k$\times$2k CCDs,
with 13.5$\mu$m ($0.33''$) pixels. Repeat exposures of 120s in $UVI$
were taken on 2002 February 2. The supernova core saturated in these
frames, and shorter 2-10s exposures were taken with the telescope
guiding continuously
between the short and long exposures to determine an accurate
position for SN~2002ap. The second set of images is from the KPNO
0.9m with the Direct Imaging Camera taken on 1993 September 15 \& 17.  
Multiple individual frames of exposure length between 400-600s
were stacked together to give total exposure
times of 5400s (in $B$), 3600s ($V$), 3200s ($R$), 
and 6900s \citep[H$\alpha$; presented previously in][]{ferg98}. 
This camera has a $2048\times2048$ TEK CCD with
$0.68''$\,pixels and the seeing in both cases was $\sim 1.5''$.  
A $K'-$band image of M74 was taken by R.\,S. de Jong using the Bok 2.3\,m.
telescope of the Steward Observatory on 1999 October 18 with the PISCES
camera \citep{mccarthy2001}; a HAWAII HgCdTe
array of 1024$\times$1024 pixels of $0.5''$ on the sky.  These
observations were mosaiced to cover the full optical disk of M74 and  
total on-source exposure time at the position of SN2002ap was 675s.

The centroid of SN~2002ap, measured in the 2002 February WFC images,
was located on all prediscovery images. Between 7 and 14 bright
stars in each of the images were used to define a
transformation with standard techniques within $IRAF$, using aperture
photometry to determine the centroids of the stars.  
The supernova position is marked 
in Fig.\,1. The errors in the SN position were calculated by taking
the quadratic sum of the transformation error and the positional error
of the supernova centroid, and are in the range $0.14-0.29''$ for $UBVRI
{\rm H}\alpha K'$. There is a nearby bright object clearly detected in $BVRI$
but, at a distance of $2.31''$ from the supernova centroid, it is
definitely not coincident with the explosion. We have estimated
limiting sensitivities for each image based on Poisson statistics 
and the 3$\sigma$ limiting magnitudes are listed in Table\,1 for each
filter. The photometric zero-points for the $UBVRI$ images were
calibrated from the photometric sequence in the field of M74 published
by \citet{henden2002}. The calibration of the $K'-$band is described
in \citet{mccarthy2001}.

In order to convert these observational limits to more useful limits
on the (de-reddened) absolute magnitudes, the extinction towards the
progenitor and distance to the host galaxy are required.
\citet{klose02} have measured the interstellar medium Na\,{\sc i} D1
absorption feature due to the gas in M74, and have determined a value
of $E(B-V)=0.008 \pm 0.002$ for the host galaxy.  Assuming a ratio of
total-to-selective extinction of 3.1, $A_V(host)=0.025 \pm 0.006$ magnitudes.
The Galactic extinction estimates are listed in Table\,1, (from Schlegel
et al. 1998, assuming R=3.1, and conversion factors from Cardelli, Clayton
\& Mathis 1989), and these clearly dominate total extinction (Galactic +
host).  There is no Cepheid distance determination to M74, and
\citet{sharina96} point out a discrepancy between two distance modulus
measurements of nearly 5 magnitudes. The distance of 7.3\,Mpc derived
by both \citet{sharina96} and \citet{sohn96} 
is intermediate between these other two, and we
assume this distance in the rest of the paper. We used the distance 
to M74 itself rather than the mean distance to the M74 group
derived by \citet{sharina96} which was used by \citet{mazz2002}. 
The difference is 0.2$^{m}$, and does not have significant consquences
for the conclusions presented below.

The detection limits in each filter can be converted to upper limits
on the bolometric luminosity of the progenitor star (as in Smartt et
al. 2001, 2002). The simple equation is

\begin{equation}
\label{eqn1}
\log L/L_{\odot} = (M_{\odot {\rm bol}} - 5 + 5 \log d + A_V - V - {\rm BC})/2.5 \\
\end{equation}

By applying the bolometric correction (BC) we obtain the upper limit for
the bolometric luminosity. However, the spectral type of the
progenitor is unknown.  To determine $\log L/L_{\odot}$ for
supergiants in the temperature range O5 to M5, the BC for each
spectral type is taken from \citet{dril2000}.  The other $UBRI$ 
filters can be used in a similar way (as
discussed in Smartt et al. 2002) to provide further constraints on
$\log L/L_{\odot}$.  The limiting values of $\log L/L_{\odot}$ as a
function of stellar effective temperature are plotted in Fig.\,2.




\section{Discussion}

SN~2002ap lies at $4'38''$ from the centre of M74. Assuming a distance
of 7.3\,Mpc, this corresponds to 9.8\,kpc, which is outside the main
area of current star formation activity. Several authors have studied
the radial abundance gradients across this galaxy from H\,{\sc ii}
region analysis \citep{mccall85, belley92, ferguson98, vanzee98}. 
From these four studies, the mean
oxygen abundance at a distance of 9.8\,kpc is $12 + \log {\rm O/H} =
8.5 \pm0.2$\,dex, which is a factor of 2 below that of solar
neighbourhood H\,{\sc ii} regions and young stars. To compare the
evolutionary states of massive evolved stars with the luminosity
limits derived, we use the stellar evolutionary tracks of
\citet{mey94} which have a metallicity $Z=0.008$. However, given the
scatter in the abundance measurements, an initial progenitor
metallicity close to solar cannot be ruled out.


In Fig.\,2 we plot on an HR diagram the upper limits on the bolometric
luminosities, together with the stellar evolutionary tracks for a
range of MS masses.  Hence there are types of massive stars at particular
evolutionary states that we can firmly rule out as being the
progenitor of this Type\,Ic supernova. These are as follows
\begin{enumerate}
\item Massive stars with initial masses greater than $\sim$30\msol\
which have evolved off the main-sequence but not yet reached the 
Wolf-Rayet stage; such as Luminous Blue Variables (LBV), and yellow
hypergiants. These stars still 
have hydrogen rich envelopes, and so our 
constraint is in agreement with the lack of hydrogen seen in 
SN~2002ap and Type\,Ic SNe in general.  We certainly would have 
detected LBVs similar to Galactic examples such as 
$\eta$Carinae, P\,Cygni, AG\,Carinae \citep{humph94}. 
\item Massive blue supergiants. The $B-$band
image should have detected a star like Sk$-69^{\circ}202$ (progenitor
of SN~1987A) at the 3$\sigma$ level. We would
certainly have detected more massive counterparts while they were
B-type supergiants.  We would have detected all B and A-type
supergiants with initial masses greater than $\sim$25\msol.
\item Red and yellow supergiants with masses greater than
approximately 15\msol. Such progenitors would have been detected in the $VRI$
passbands.  This is consistent with the suggestion of \citet{smartt02}
that normal SN II-P may come from moderate mass progenitors
(M$\lesssim$12\msol) in the red supergiant phase.
\end{enumerate}

We cannot dismiss lower mass progenitors (less than approximately
15\msol) at any stage of their evolutionary lives.  However {\em
single} stars in this mass range are unlikely to lose their hydrogen
atmospheres before they reach the end of their lives, and so it is
virtually inconceivable that they produce SNe~Ic.  The prediscovery
images are not sensitive to all typical magnitudes of WR stars. We
stopped the shaded region in Fig.\,2 at the edge of the O-star
main-sequence (ZAMS mass of 85\msol\ and $T_{\rm eff}$=48000K), 
to consider the WR region separately. 
Wolf-Rayet stars span a large range in absolute
magnitudes. For example, in the Galaxy and the LMC they have continuum
magnitudes in the range $-8 \lesssim M_{b} \lesssim -3$
although it is only rarely that they have magnitudes
at the brighter end of this range \citep{vacca90}.  However these
magnitudes are measured with narrow-band filters to sample
continuum regions free from the characteristic 
broad, strong emission lines\footnote{The $ubv$ filters often used are
typically 100\AA\ wide and centred near 3650\AA, 4270\AA\ and 5160\AA\
respectively}. Magnitude differences $b-B$ and $v-V$ can be
up to $-0.55^{m}$ and  $-0.75^{m}$ respectively 
for WC stars i.e. the strongest
line WC stars could be $0.55^{m}$ brighter in $B$ than the
\citet{vacca90} $b$ magnitudes. Taking that sample as representative,
our sensitivity limit of $M_{B}=-6.3$ should permit the detection of
roughly 30\% of WR stars. Assuming a $BC=-4.5$ is appropriate for WR
stars \citep{crow2002,smith89}, then 
Eqn.\,1 suggests an upper limit to the
luminosity of a WR star progenitor of $\log L/L_{\odot} \lesssim 6.2$.
Applying the approximate mass-luminosity relation for WR stars from
\citet{maeder83}, this corresponds to an initial upper mass limit of
$\sim$40\msol. These approximate numbers rule out a very high mass WR
progenitor, but about 70\% of typical WR stars (of initial masses less than
$\sim$40\msol) would not be detected and so are viable progenitors.

\citet{mazz2002} have presented a preliminary model of the early
evolution  of SN~2002ap, finding a kinetic
energy of $\sim4-10 \times 10^{51}$\,ergs and an ejected heavy element
mass of M$_{\rm ej} \simeq 2.5-5$\msol.  They suggest that this is most
consistent with an explosion of a $\sim$5\msol\ C+O star, which would
have had an initial main-sequence mass of M$_{\rm ms} \sim
20-25$\msol.  The explosion was less energetic than that of SNe~1998bw
or 1997ef \citep[$\sim5 \times 10^{52}$\,ergs;][]{nakamura01} although
the very broad spectral features indicate that SN~2002ap is of similar
nature to these hypernovae. Our non-detection forces us to conclude
that a M$_{\rm ms} \sim 20-25$\msol\ progenitor star must either have
been in an evolutionary phase 
hotter than $\sim$15000K (or else we would have detected it on
the prediscovery images - see Fig.\,2), 
or the star did not go through classical single star
evolution. Given that the progenitor must have lost its hydrogen
envelope to become a Type~Ic, a M$_{\rm ms} \sim 20-25$\msol\
progenitor in an interacting binary system appears to be the most
consistent explanation for the progenitor non-detection
and the C+O core mass inferred in the
\citet{mazz2002} analysis. As discussed above, the prediscovery images
would not be sensitive to the majority of WR stars, and on this basis
alone we cannot rule them out. At LMC type
metallicity, WR stars should have initial masses of $\ge$30\msol\
\citep{massey2000}, and final C+O cores of $\sim$10\msol. 
Although this appears a factor of two higher than inferred from the
the \citet{mazz2002} explosion models, the final 
masses of WR stars are somewhat uncertain, and could be as low as
7\msol\ \citep{dray2002,crow2002}. 
Also as noted above, the metallicity of the 2002ap region we have 
adopted is not definitive and {\em could} be close to solar, which would
result in a lower initial mass for WR formation, and a slight lowering of
the core mass. Given all of these uncertainties
we cannot distinguish between the single WR scenario and 
the progenitor being a $20-25$\msol\ star
in an interacting binary system which has had its outer H-rich
envelope removed due to mass transfer.  The latter scenario
was first theoretically suggested by \citet{nomoto95} and the observational
work of \citet{vandyk96} on the association of Type II, Ib and Ic 
events with massive H{\sc ii} regions supports the idea that the
progenitors of SNe Ib and Ic could be 
interacting binary stars, rather than initially very massive stars
which have reached the WR phase.  
The results of this paper together
with those of \citet{mazz2002} may suggest a similar origin for SN~2002ap.
In comparison, hydrodynamic modelling of the
lightcurve and spectra of SN~1998bw suggests the explosion of a
14\msol\ C+O star which is the core of a star of initial mass 40\msol\
\citep{nakamura01}. This could certainly be a WR star, although
SN~1998bw was too distant to allow constraints on
its progenitor from prediscovery images. 

The position of SN~2002ap is $2.31''$ (or 80\,pc) away from the bright
object clearly detected in the $BVRI$ images shown in Fig.\,1, with only a
a faint sign of the object in the $U-$band. At H$\alpha$
there is weak, possibly extended emission although the SN position
is clearly not coincident with any nebular flux.  
This object has magnitudes (again measured with respect
to the calibration of Henden 2002) $B=21.50, B-V=0.44, V-R=0.19,
R-I=0.37$. This source shows some evidence of being more extended than
the typical point-spread-function of the image, although the
resolution and variable background are such that much higher
resolution images are required to determine its exact nature. 
Assuming a similar extinction to
this object as for the supernova, it has $M_B = -7.8$, and colours
which would be consistent with it being a very luminous {\em single}
supergiant star of type late A or F with a mass roughly of $\sim$40\msol. 
Alternatively it could be an
unresolved cluster with a diameter similar to, or less than, the
seeing disk i.e. roughly 100\,pc.  It is unlikely that massive stars
form individually, and one would expect that if a WR star or
20-25\msol\ initial mass binary component was the progenitor
then an accompanying population of stars of lower or 
equivalent mass should be seen. The absolute $B$-magnitude
of this object is roughly consistent with a star cluster with a mass
 $5^{+5}_{-3} \times 10^{3}$\msol\ \citep[from the Galaxy
Isochrone Synthesis Spectral Evolution Library (GISSEL) of][]{bru93},
although its colours are somewhat redder than one would expect from a
young cluster of age less than $\sim$100\,Myrs.  The depth of the
$K$-band is rather too shallow to constrain individual stellar
objects. However the lack of any significant flux shows that there is
no large scale starforming region enshrouded in dust, which could be
host to an optically hidden large population of massive stars.  Once
SN~2002ap has faded significantly, it is imperative that deep, high
resolution images are taken of its environment.  Ideally this should
be done with the {\em Hubble Space Telescope}.  With the {\em Advanced
Camera for Surveys} one could 
resolve the nature of this object, and 
construct an accurate CMD (down to $M_V\sim-2$ in a very modest
amount of time) to constrain the star formation
history of this region. If it remains a single object then detection of 
lower mass stars in this region will still produce 
an extinction map from multi-colour photometry. 
Furthermore, if the progenitor was part of an 
interacting binary system, it is likely that the companion is a
fairly massive star and may be detectable in deep images several
years from now.

In summary, the prediscovery images allow significant
constraints to be placed on the nature of the progenitor of the
nearest Type\,Ic supernova (and probable hypernova) to have occurred
in modern times. We can rule out various evolutionary states of
massive stars which would be clearly detectable on the pre-explosion
images.  These include very high mass ($\gtrsim 40$\msol) WR stars,
although this still leaves roughly 70\% of typical WR-types as viable
progenitors.  
We cannot distinguish between
the WR model and  the death of a star with initial mass
$\sim$20-25\msol\ which has had its outer envelope stripped off
through mass transfer in a binary system. 
However, this unexcluded fraction of WR stars 
may have too high a final core mass to be consistent with initial 
models of the supernova energetics. The galaxy
M74 has been imaged by HST, Gemini, CFHT and WHT, however the
supernova position does not fall on any of these images.  We have
searched all the publicly available archives for deeper, higher
resolution images of M74 but have found no superior images to those
presented here that include the pre-explosion site of SN~2002ap.



\acknowledgments
SJS acknowledges support from PPARC, and
ERR acknowledges CONACYT, SEP and the ORS foundation. 
The data were made publically available through the
Isaac Newton Groups' Wide Field Camera Survey
Programme. The Isaac Newton Telescope is operated
on the island of La Palma by the Isaac Newton
Group in the Spanish Observatorio del Roque de los
Muchachos. We acknowledge access to the Wide Field Survey's 
data products from the 
Cambridge Astronomical Survey Unit at the Institute of Astronomy. We thank
R. de Jong for the $K'-$band image, and P. Crowther and 
P. Royer for advice on WR stars.






\begin{table}
\begin{center}
\caption{The 3$\sigma$ limiting magnitudes derived from pre-discovery
images obtained at the Isaac Newton and KPNO\,0.9m telescopes
($UBVRI$), and the Bok 2.3m telescope ($K'$). The absolute magnitude
limits are calculated assuming a distance of 7.3\,Mpc with extinctions
as described in Section\,2. In each band the value quoted uses the
deepest limit available.}
\begin{tabular}{llccc}
\tableline\tableline
Filter  & INT   & KPNO 0.9m  & A$_{\lambda}$ &  Abs Mag.\\ \tableline
U       & 21.5  &  ...   & 0.387      &  $-$8.2\\
B       & 22.7  &  23.3  & 0.307      &  $-$6.3\\
V       & 22.6  &  22.9  & 0.236      &  $-$6.6\\
R       &  ...  &  22.2  & 0.190      &  $-$7.3\\
I       & 21.5  &  ...   & 0.138      &  $-$7.9\\
K       & 18.1  &  ...   & 0.026      &  $-$11.2\\
\tableline						  
\end{tabular}	
\end{center}
\end{table}
\clearpage

\begin{figure}
\epsscale{0.8}
\plotone{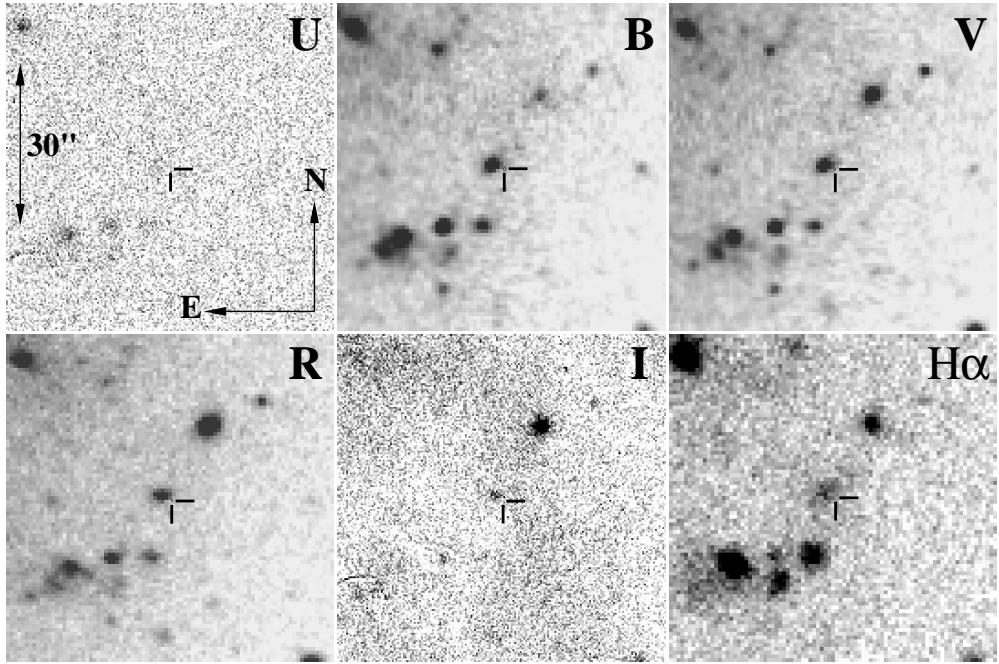}
\caption{The prediscovery optical images, with $BVR{\rm H}\alpha$ from the
KPNO\,0.9m and $UI$ from the INT WFC. The location of SN~2002ap is at
the centre of each frame, indicated by the orthogonal lines. The SN
position is $2.31'' \pm 0.29''$ away from the nearby bright object
detected in $BVRI$ (and marginally seen in $U$) i.e. it is clearly not
coincident with this source.}
\end{figure}
\clearpage

\begin{figure}
\epsscale{1.0}
\plotone{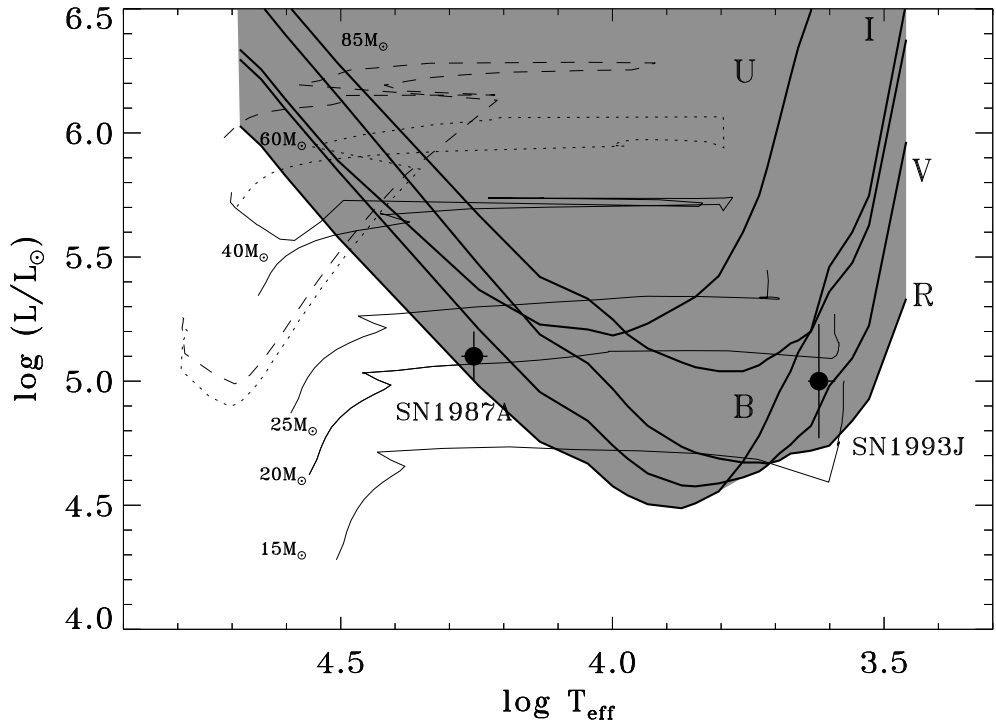}
\caption{The Geneva evolutionary tracks \citep{mey94,sch92} for
15$-$85\,M$_{\odot}$ plotted on an HR diagram.  The positions of the
progenitors of SN~1987A and SN~1993J are indicated.  The luminosity
limits as a function of stellar effective temperature are plotted as
the thick solid lines. The prediscovery images are sensitive to all
stars lying in the shaded regions of the HR diagram.}
\end{figure}
\clearpage

\end{document}